# Lateral Ventricular Brain-Computer Interface System with Lantern-Inspired Electrode for Stable Performance and Memory Decoding

Yike Sun[1,2,+], Yaxuan Gao[3,+], Kewei Wang[1,+], Jingnan Sun[1], Yuzhen Chen[1], Yanan Yang[3], Tianhua Zhao[3], Haochen Zhu[3], Ran Liu[1,*], Xiaogang Chen[4,*], Bai Lu[3,*] and Xiaorong Gao[1,*]

+Y. S., Y. G. and K. W. contributed equally to this work.

1 School of Biomedical Engineering, Tsinghua University, Beijing, 100084, China.

2 School of Biological Science and Medical Engineering, Beihang University, Beijing, 100191, China.

3 School of Pharmaceutical Sciences, Tsinghua University, Beijing, 100084, China.

4 Institute of Biomedical Engineering, Chinese Academy of Medical Sciences and Peking Union Medical College, Tianjin, 300192, China.

*Correspondence address. Tsinghua University, Beijing, 100084, China.

Email: liuran@tsinghua.edu.cn; chenxg@bme.cams.cn; lubailab@gmail.com; gxr-dea@mail.tsinghua.edu.cn.

# Abstract

We present a lateral ventricular brain-computer interface (LV-BCI) that deploys an expandable, flexible electrode into the lateral ventricle through a minimally invasive external ventricular drainage pathway. Inspired by the framework of traditional Chinese lanterns, the electrode expands uniformly within the ventricle and conforms to the ependymal wall. Compared with conventional subdural ECoG electrodes, the LV-BCI shows superior signal stability and immunocompatibility. Resting-state spectral analyses revealed a maximum effective bandwidth comparable to subdural ECoG. In evoked potential tests, the LV-BCI maintained a consistently higher signal-to-noise ratio over 112 days without the decline typically associated with scarring or other immune responses. Immunohistochemistry showed only a transient, early microglial activation after implantation, returning to control levels and remaining stable through 168 days. We further designed an "action-memory T-maze" task and developed a microstate sequence classifier (MSSC) to predict rats' left/right turn decisions. The LV-BCI achieved prediction accuracy up to 98%, significantly outperforming subdural ECoG, indicating enhanced access to decision-related information from deep structures such as the hippocampus. These results establish the lateral ventricle as a viable route for neural signal acquisition. Using a lantern-inspired flexible electrode, we achieve long-term stable recordings and robust memory decision decoding from within the ventricular system, opening new directions for BCI technology and systems neuroscience.

# Introduction

Understanding consciousness and brain function remains at the forefront of science. Advances in neuroscience and engineering are steadily deepening our insight into this complex system[1]. Brain-computer interfaces (BCIs) offer a direct pathway to read and modulate neural activity, enabling new therapeutic and assistive applications[2,3]. A central challenge, however, is achieving durable-neural signal acquisition[4]. While most BCIs to date rely on non-invasive recordings, skull and tissue attenuation limit signal quality and spatial resolution. Implantable approaches promise higher fidelity. Prevailing methods, such as electrocorticography (ECoG) [5] and microelectrode arrays (MEAs) [6], place devices on or within neural tissue. However, the mechanical and geometric mismatches between devices and soft brain tissue can induce damage and immune responses[7], compromising safety and long-term performance.

Endovascular stent-electrode strategies have recently emerged as minimally invasive alternatives[8,9], leveraging vascular spaces to achieve stable long-term recording and have shown clinical feasibility[10]. Yet they carry intrinsic risks of thrombosis and in-stent restenosis, often requiring long-term antithrombotic therapy[11]. Moreover, small vessel calibers and complex anatomy constrain device size and coverage, limiting bandwidth and channel placement. These constraints motivate exploration of alternative anatomical routes that can support long-term, wide-bandwidth neural recording without chronic vascular risks.

We focus on the lateral ventricles (LV), the largest natural fluid-filled cavities in the brain. Cerebrospinal fluid (CSF) offers a sparsely cellular, mechanically gentle, and stable milieu compared with blood vessels[12]; CSF circulation is slow and contains virtually no red blood cells, reducing thrombotic risk[13]. Although ventricular spaces are deep, developmental and anatomical evidence indicates close relationships between ventricular zones and cortical structures[14,15], supporting the feasibility of recording cortical signals from ventricular walls[14-16]. Early studies have demonstrated BCI-relevant potentials from the ventricular system[17], and ventricular implants are technically feasible[18]. We hypothesize that the lateral ventricle—at the core of the ventricular system—provides an even more stable physicochemical environment and is closer to deep structures such as the hippocampus and thalamus, offering unique opportunities for BCI paradigms targeting memory and decision-making.

Clinically, ventricular access is routine in procedures such as external ventricular drainage, providing an established surgical pathway for device delivery[19]. By combining this approach with an expandable electrode, a minimally invasive and effective BCI interface can be achieved. However, because the ependymal wall is less resilient than vascular walls, rigid stent-like devices risk tissue injury. Addressing this risk requires the development of new implantation techniques as well as soft, compliant electrodes.

Here, we introduce a lateral ventricular BCI (LV-BCI) that enables long-term stable neural recording and high BCI performance. The system employs an expandable flexible electrode delivered via an external ventricular drainage path and deployed within the lateral ventricle. Compared with conventional subdural ECoG, the LV-BCI provides more stable long-term signal quality with reduced immune response. Leveraging the ventricle's proximity to the hippocampus, we decode memory-driven decision processes in rats—a task that is challenging for standard cortical BCIs. By exploiting the unique advantages of the ventricular system with a lantern-inspired flexible design, the LV-BCI offers a promising route to overcome limitations of current BCI technologies and opens new avenues for systems neuroscience.

## Results

**Lantern-inspired flexible ventricular implant electrode**

The limited elasticity of the ependymal lining of the lateral ventricle makes rigid, metal stent-like electrodes prone to cutting injury and are therefore unsuitable for chronic implantation. Drawing inspiration from the framework of traditional Chinese lanterns, we designed a new expandable, flexible electrode (Fig. 1a). In lanterns, a central cylinder bears the primary load while a compliant outer lattice collapses and expands (Fig.

1b, left). Analogously, our electrode follows this skeletal layout (Fig. 1b, right): it is compact when unloaded and expands under gentle force. Upon deployment in the ventricle, bow-shaped side contacts spring outward to conform closely to the ventricular wall, ensuring intimate contact. Notably, this radially symmetric structure may help dissipate shear forces arising from CSF pulsation, reducing micromotion and improving stability (hypothesis). Surgically, we adapted a standard external ventricular drainage (EVD) trajectory[20]. Using a stereotaxic frame, the folded electrode is inserted through a small cranial opening and advanced to the target, then expanded in situ (Fig. 1c), minimizing cortical exposure and parenchymal disruption.

The device integrates flexible electronics with programmable micro-assembly. We first fabricated bendable flexible printed circuit board (fPCB) electrodes, then laser-cut and reconfigured them into a three-dimensional lantern form using an origami-inspired process (see Methods). Each contact adopts a three-dimensional bow-arch geometry that improves stress distribution, akin to arch bridges, leveraging advances in kirigami/origami-inspired bioelectronics. To evaluate mechanical compatibility in the ventricular environment, we built finite element models in ANSYS Workbench comparing the lantern electrode with a conventional metal stent under matched displacement loading (Fig. 1d; parameters in Supplementary Table 1). Under a 0.001 mm imposed displacement, the lantern's arch skeleton absorbed most strain energy, limiting peak equivalent deformation at the ventricular wall to 0.001 mm This demonstrates that the electrode's elastic deformation effectively dissipates the load. In contrast, the metal stent induced 0.0015 mm peak wall deformation (Supplementary Fig. 1), a 33% reduction with the lantern topology, suggesting effective mitigation of shear stress concentration at the tissue interface.

We further tested structural integrity during deployment using an explicit dynamic, multi-material finite element model (Supplementary Table 2). Dynamic simulations (Fig. 1e) showed that during unfolding from the compact state to full expansion (0.8 s), the copper interconnects experienced peak equivalent stress <15 MPa (3% of copper yield strength), and the polyimide substrate <8 MPa (12% of its tensile strength). Local stress concentrated at gold layer edges, with a transient peak 150 MPa, near the gold yield range (120-150

MPa). Johnson-Cook plasticity modeling indicated 0.2% plastic strain at those hotspots, well below microcrack initiation thresholds (>1.5%), indicating no structural failure during deployment. These simulation results support the electrode's favorable mechanics during unfolding and wall apposition and suggest that using even more ductile surface materials could further enhance damage tolerance.

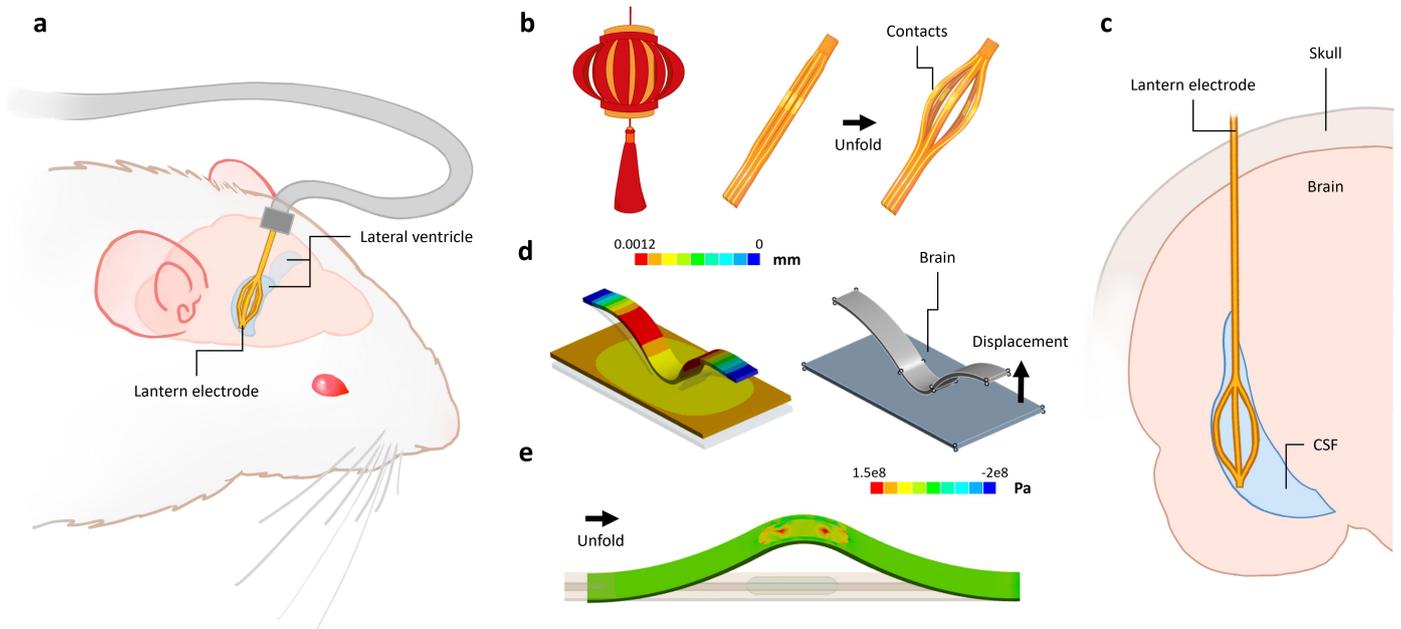

**Figure 1.** Lantern-inspired flexible ventricular electrode. a, Schematic of the minimally invasive LV-BCI design. b, Traditional lantern framework (left) and the derived 3D flexible electrode structure with deployment sequence (right). c, Schematic of implantation via an EVD-like path. d, Finite element analysis of a lantern contact (left) and model schematic (right); displacements in millimeters, arrows indicate direction. e, Explicit dynamics simulation of unfolding within the lateral ventricle.

**Long-term stable evoked-potential recordings**

Most non-cortical implant BCI studies focus on extracting and decoding features of event-related potentials (ERPs) [21,22]. To evaluate our minimally invasive LV-BCI system, we used conventional subdural (pial surface) planar ECoG as a control. Because the ventricular ependyma interposes between the LV electrode and brain tissue and is biophysically similar to the pia for signal transmission, this control ensures a fair comparison. To exclude material-related confounds, we fabricated a second control using the same fPCB material in a planar form (no origami) and placed it on the cortical surface above the left lateral ventricle in rats (implant sites in Fig. 2a). We first analyzed resting-state power spectra and computed the maximum effective bandwidth following Oxley et al. [8]. Both LV and ECoG groups showed comparable bandwidths fluctuating around 380 Hz within 112 days post-implantation, indicating that ventricular recordings achieve a maximum effective bandwidth on par with subdural ECoG (Fig. 2b,c).

We next compared ERPs using steady-state visual evoked potentials (SSVEPs). The stimulation hardware, visual protocol, and software platform are described in Methods. As shown in Fig. 2e, anesthetized rats were positioned in a dark chamber facing a stimulus screen, and we recorded multi-trial, multi-frequency SSVEPs (timeline in Fig. 2d). At postoperative day 7 with 14 Hz stimulation, both groups exhibited modest amplitudes due to anesthesia, but the ECoG group showed stronger SSVEP peaks (Fig. 2f). Across 112 days, narrowband SNR (Fig. 2g) initially favored ECoG at day 7 (ECoG −7.26±0.24 dB vs LV -8.15±0.06 dB; $p=0.04<0.05$), but then declined in ECoG with a slope of −0.19 dB, while LV showed no clear decay. By day 56, LV surpassed ECoG significantly (LV −8.10±0.04 dB vs ECoG −8.68±0.09 dB; $p=0.01<0.05$) and maintained a >0.5 dB advantage thereafter. Although not every session reached statistical significance, the overall trend indicates a slower SNR decay in the LV group. Given identical electrode materials, this advantage likely stems from the more stable ventricular environment and reduced immune response.

Building on the visual experiments, we conducted auditory steady-state response (ASSR) tests. The auditory cortex lies relatively deep and is covered by complex musculature, making high-quality recordings challenging in clinical Electroencephalogram (EEG). The LV electrode, approaching the auditory cortex from the ventricular side, may provide a new recording route. Using a multi-trial, multi-frequency protocol (Fig. 2h; timing in Fig. 2d), we observed that at day 7 with 40 Hz stimulation the LV group exhibited stronger overall narrowband SNR, especially at the fundamental and first two harmonics (Fig. 2i). Over time (Fig. 2j), LV maintained a significant SNR advantage from day 7 to day 112 (all timepoints $p<0.05$). On day 7, LV averaged −5.37±0.20 dB versus ECoG −6.75±0.30 dB ($p=0.04<0.05$). By day 112, LV remained stable at −5.54±0.28 dB while ECoG declined to −9.17±0.43 dB (p=0.002<0.01), demonstrating superior long-term fidelity for LV.

Interestingly, despite its lower overall SNR, the ECoG group showed relatively higher SNR at higher-order harmonics (e.g., the fourth). Fig. 2k summarizes harmonic-wise SNRs across timepoints; Fig. 2l and 2m track first- and fourth-harmonic SNR trajectories, respectively. Both groups declined over time, but ECoG maintained a relative advantage at the fourth harmonic (Fig. 2m). This may reflect different network sampling: cortical-surface electrodes are closer to higher-order auditory processing and capture more nonlinear harmonic content, whereas ventricular electrodes are nearer primary auditory circuits and are more sensitive to the fundamental. These results suggest that the lantern-inspired electrode can complement traditional cortical arrays, enabling multi-level interrogation of sensory pathways.

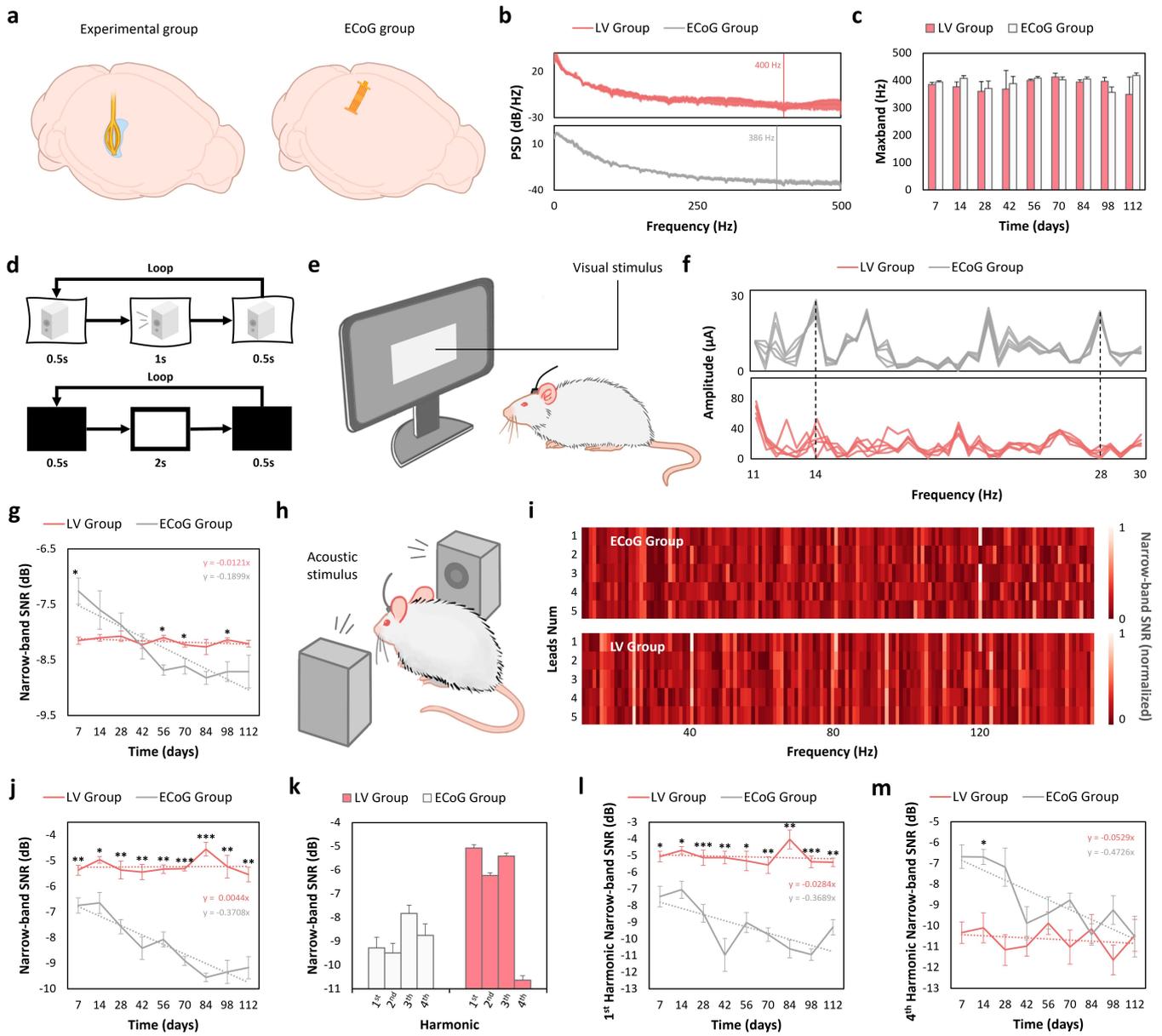

**Figure 2.** Comparison of evoked potentials between LV flexible electrodes and cortical ECoG. a, Implant locations. b, Resting-state PSDs with vertical lines marking maximum effective bandwidth. c, Time course of maximum effective bandwidth. d, SSVEP and ASSR timelines. e, SSVEP setup. f, Power spectra at 14 Hz on day 7 (trial-averaged). g, Narrowband SNR over time for SSVEP. h, ASSR setup. i, Narrowband SNR heatmaps at 40 Hz on day 7. j, Narrowband SNR over time for ASSR. k, Harmonic-wise SNRs at each timepoint. l, First-harmonic SNR over time. m, Fourth-harmonic SNR over time. Error bars denote ±SEM. Asterisks indicate statistical significance (*P<0.05; **P<0.01; ***P<0.001).

**Biocompatibility during chronic implantation**

The ERP experiments above showed that, compared with conventional subdural ECoG, the lantern-inspired electrode enables more stable long-term neural recording. We hypothesized that this advantage arises from lower immune activation at the electrode-tissue interface. Using SD rats, we performed comparative immunohistochemistry to assess how different implantation routes shape the neuroimmune milieu over time.

Fig. 3a illustrates representative immunostaining at three key time points—postoperative weeks 2, 12, and 24—for the subdural ECoG group, the LV group, and sham controls. In the ECoG group, robust immune activation was evident around the implant-tissue interface, with clusters of microglia showing activated morphology (enlarged somata with retracted, thickened processes), in contrast to the sparser cells in controls. In the ventricular group, staining patterns were largely similar to controls, suggesting limited parenchymal immune activation; however, a transient microglial aggregation was observed near the superior ventricular wall at 2 weeks, which rapidly subsided thereafter, consistent with an acute response to surgical entry.

Because the targeted brain regions differed across groups (cortex for ECoG vs ventricle for LV), conventional intensity-based metrics are not ideal for cross-region comparisons. We therefore established quantitative indices based on the microglia-to-neuron ratio and microglial soma size (Fig. 3b), counting cells within anatomically matched regions (Fig. 3c).

In the subdural ECoG group, microglia/neuron ratios at 2, 12, and 24 weeks were 0.822, 0.766, and 0.761, respectively, versus 0.401, 0.284, and 0.236 in controls (P=0.007, 0.004, 0.002; Fig. 3d), indicating persistent immune activation. In contrast, the ventricular group showed a significant elevation only at 2 weeks (0.734 vs control 0.401, P=0.033). By 12 and 24 weeks, ratios declined to 0.342 and 0.385, not significantly different from time-matched controls (0.407 and 0.332; both P>0.5; Fig. 3e), indicating that activation was confined to the early postoperative period.

Microglial soma areas (Fig. 3f, g) in the ECoG group measured 61.9 μm² and 54.2 μm² at weeks 2 and 12,

slightly larger than controls (39.0 μm² and 36.2 μm²) but not statistically significant (P=0.067, 0.173). At week 24, soma area increased to 56.9 μm² versus 33.7 μm² in controls (P=0.049; Fig. 3f), reflecting sustained activation-associated morphological changes. In the ventricular group, soma area was significantly enlarged at 2 weeks (60.0 μm² vs 35.4 μm²; P=0.026) but returned to control-like levels by weeks 12 and 24 (Fig. 3g), consistent with a transient acute response.

Together, these data indicate that subdural ECoG produces sustained and pronounced microglial activation—higher microglial density and enlarged somata—whereas the lantern-inspired electrode induces only an early, one-off response that rapidly diminishes and shows no long-term difference from controls. Given that both electrodes share the same materials, the divergence likely reflects implantation site and mechanical compliance: the ventricular device resides within the CSF pathway, where its flexible structure reduces friction with parenchyma, and CSF circulation may facilitate clearance of inflammatory mediators, promoting resolution of immune activation[13,23,24]. Notably, although statistical differences disappeared after 12 weeks in the ventricular group, absolute values remained slightly above controls, suggesting a weak, residual stimulus may persist with chronic implantation.

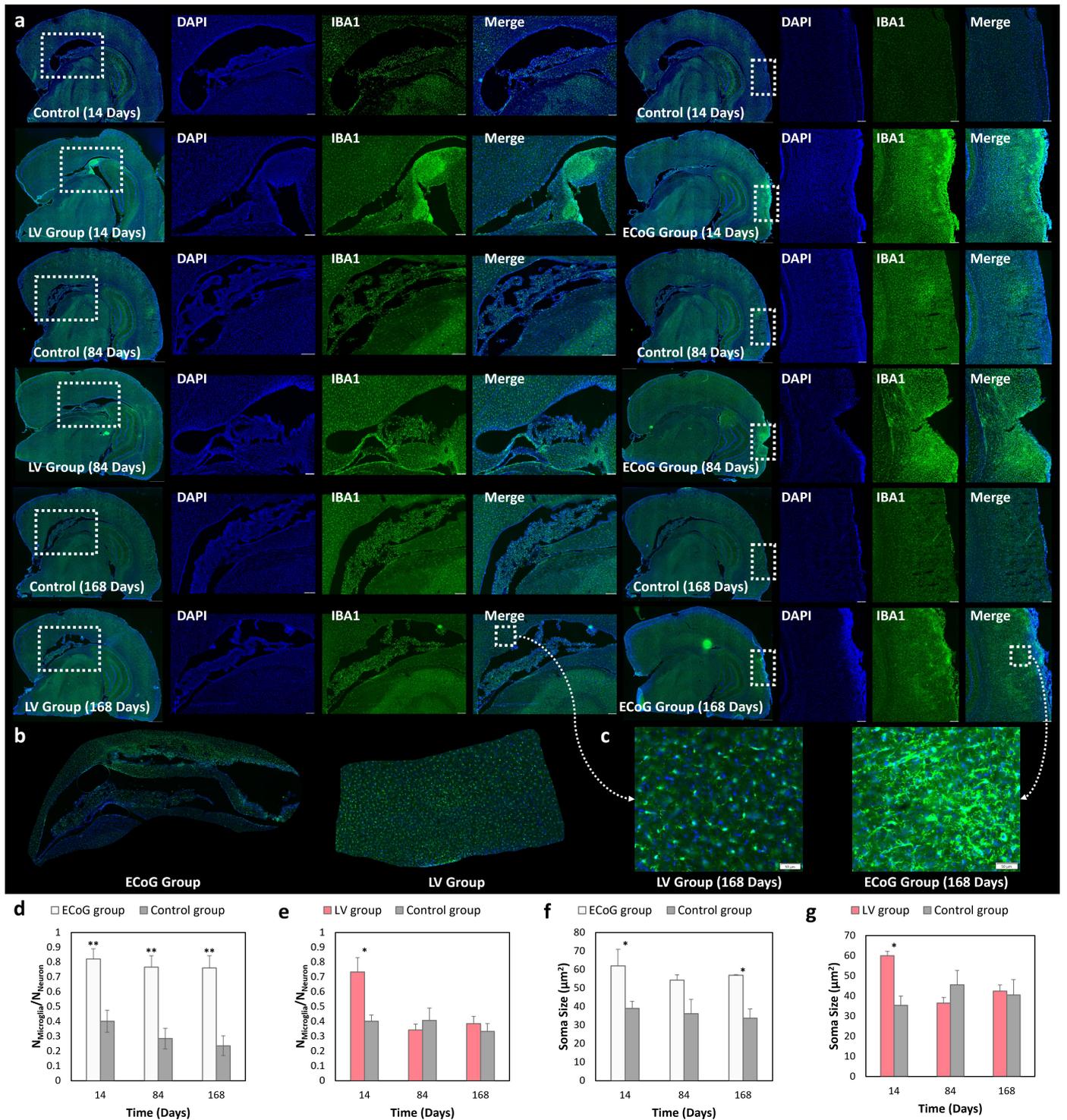

**Figure 3.** Immunological analysis. a, Representative immunostaining at 14-, 84-, and 168-days post-surgery across groups (sham, subdural ECoG, lantern-inspired electrode). b, Regions of interest for quantification. c, Cell morphology after 168 days of implantation. d, Time courses of microglia/neuron ratio for ECoG vs control. e, Time courses for ventricular vs control. f, Microglial soma area over time for ECoG vs control. g, Soma area over time for ventricular vs control. Error bars denote ±SEM. Asterisks indicate statistical significance (*P<0.05; **P<0.01; ***P<0.001).

**Memory decision decoding task**

Because the LV electrode lies near deep structures such as the hippocampus, we designed a memory-driven decision decoding paradigm to test its BCI utility (Fig. 4a). Following prior work[25], we built an "action-memory T-maze" task. In each trial, a rat was confined in the start chamber while a cue light in the left or right arm of the T-maze was illuminated for 10 seconds and then turned off. After a 10-second delay, we recorded 30 seconds of EEG during which we aimed to predict the rat's upcoming left/right turn based on neural activity. The door of the chamber was then opened and the rat was allowed to move; correct choices (toward the previously lit arm) were rewarded (Fig. 4b).

We developed a microstate sequence classifier (MSSC) to decode memory-related decision intent (Fig. 4c), motivated by EEG microstate theory[26,27]. Because the onset of internal recall during the pre-movement window was unknown, we converted continuous EEG into a sequence of microstates and searched for discriminative patterns. Continuous signals were segmented by a sliding window; the mean absolute multi-channel amplitude within each window defined a microstate feature. Microstates from the training set were clustered using K-means, and clusters with the strongest left/right differences were selected as discriminative states; all remaining clusters were encoded as non-discriminative (label 0). Each trial's time series was then mapped to a discrete label sequence and fed into an SVM with a nonlinear kernel. Leave-one-out cross-validation showed robust separation of left versus right decisions, indicating that pre-decisional microstate sequences carry predictive information.

Classification performance is summarized in Fig. 4d for the LV and ECoG cohorts (n=5 rats per group; 20 trials per rat, balanced left/right). In the full band (>0.5 Hz), the LV group reached 88.6%±2.2%, significantly above chance (50%, P<0.001) and higher than the ECoG group at 64.0%±3.6% (P=0.0177). Band-specific analyses showed that LV decoding remained robust in the $\delta$ (0.5-4 Hz; 85.8%±4.9%), $\alpha$ (8-13 Hz; 89.6%±4.3%), and $\gamma$ (30-80 Hz; 74.6%±5.1%) ranges (all P<0.05 vs 50%) and exceeded ECoG performance in each case. By contrast, ECoG accuracies in these bands hovered near chance (δ 57.0%±3.0%, α 51.0%±2.97%, γ 51.0%±3.85%; all not significant vs 50%). In θ (4-8 Hz), β (13-30 Hz), and high-γ (>80 Hz), LV accuracies were 65.8%, 66.1%, and 53.5%, respectively; only θ slightly exceeded chance (P ≈ 0.054), and none differed significantly from ECoG (θ 62.0%, β 46.0%, high-γ 48.0%). Taken together, LV outperformed ECoG in the full band and several key sub-bands (e.g., full band and α, both P<0.01), suggesting that ventricular recordings carry richer pre-decisional information, likely reflecting contributions from hippocampal-diencephalic circuits. This interpretation is consistent with reports that hippocampal theta compresses spatial trajectories in time[28], providing a substrate for prospective path planning.

We further introduced filter-bank strategies to better exploit band-specific information, a common approach in EEG/BCI decoding[29,30]. As shown in Fig. 4e, we evaluated two schemes: method A uses non-overlapping bandpass filters with weighted fusion of MSSC outputs; method B uses overlapping high-pass filters in the high-frequency range. In the LV group (Fig. 4f), filter-bank MSSC method A yielded 90.8%±0.04%, slightly above full-band MSSC (88.6%±2.2%) without significance (P>0.05). Method B reached 98.0%±0.01%, significantly higher than full-band MSSC (P=0.009<0.01), improving both mean accuracy and reducing SEM, indicating more effective use of complementary spectral information.

Fig. 4g maps the spatial distribution of microstates selected as discriminative features under full-band MSSC. In the LV group, feature microstates concentrated in the striatum (STR) and dorsal hippocampus (HIP-D), whereas ECoG features localized mainly to parietal cortex (PTV, PTA). Given the hippocampus lies beneath parietal cortex in rats, sources may partially overlap; however, the ventricular electrode, immersed in CSF and closer to deep structures (e.g., hippocampal subfields and thalamus), likely captured more memory-decision-related activity[31].

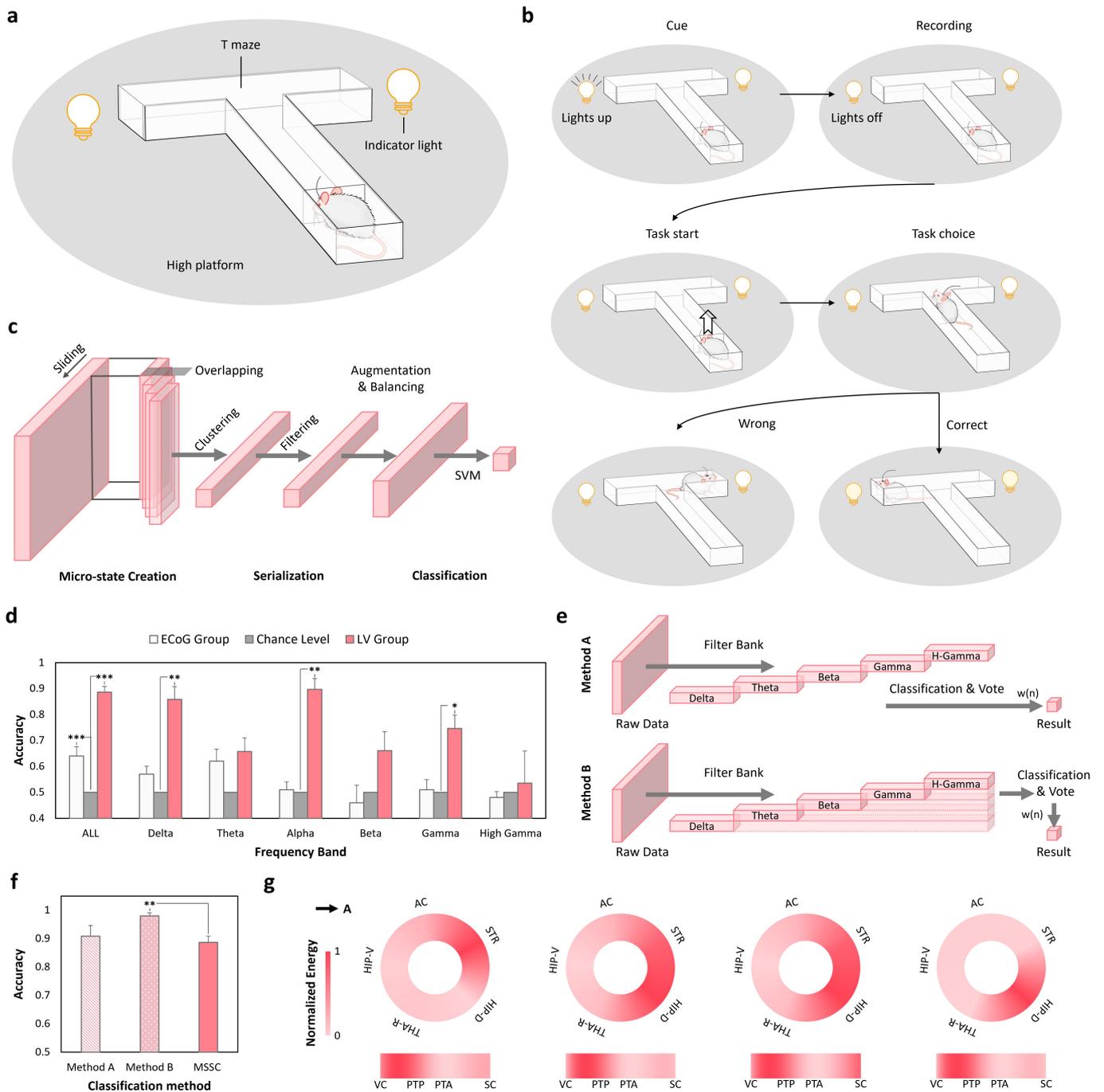

**Figure 4.** Memory decision decoding with the LV BCI. a, Schematic of the behavioral platform: a rat is restrained facing a T-maze in a dark chamber. b, Task timeline: a cue light in the left or right arm is presented, followed by a pre-decision recording window, after which the door opens for the rat to choose. c, Workflow of the microstate sequence classifier (MSSC). d, Classification accuracy of LV and ECoG groups across frequency bands. e, Two filter-bank schemes: Method A uses non-overlapping bandpass sets; Method B uses overlapping high-pass sets at the high-frequency end. f, Classification accuracy with filter-bank MSSC in the LV group. g, Spatial topography of normalized microstates selected as discriminative features

in full-band MSSC for LV and ECoG. AC, auditory cortex; STR, striatum; HIP-D, dorsal hippocampus; THA-R, thalamus; HIP-V, ventral hippocampus; VC, visual cortex; PTP, posterior parietal cortex; PTA, parietal association cortex; SC, somatosensory cortex. Error bars denote ±SEM. Asterisks indicate statistical significance (*P<0.05; **P<0.01; ***P<0.001).

**Discussion**

We introduce a LV-BCI that provides a new route for chronically stable neural recording. A lantern-inspired flexible electrode enables improved mechanical matching with brain tissue compared with conventional rigid or planar devices. Functional testing shows high long-term stability for evoked-potential recordings. Subdural electrodes, which contact mature cortical tissue directly, often trigger chronic microglial activation and scarring that degrade signal quality. By contrast, the LV-BCI resides within CSF, where neural and vascular elements are sparse; after an initial surgical insult, the immune milieu rapidly settles, and CSF circulation likely facilitates dispersion and clearance of inflammatory mediators[32,33]. Consistent with this, immunohistochemistry revealed sustained microglial activation with subdural ECoG but only a transient, early response with the ventricular electrode, which then returned to control levels.

The ventricular system remains underexplored electrophysiologically. Most EEG/ECoG studies interrogate the cortical surface, whereas the ependymal lining—the brain's inner surface—has received little attention. Developmental biology places neural progenitors along the ventricular wall and documents large-scale neuronal migration from ventricle to cortex, suggesting that a portion of cortical activity can be conducted to the ventricular boundary. Our SSVEP and ASSR results demonstrate that ventricular electrodes can capture canonical sensory-evoked signals. We also observed that ECoG sometimes exhibited stronger higher-order harmonics, consistent with its proximity to higher-order cortical processing and nonlinear network dynamics, whereas LV-BCI was more sensitive to fundamentals, indicating stronger coupling to more primary pathways. This complementarity suggests that ventricular and cortical electrodes could be combined for multi-level electrophysiological mapping.

Leveraging the lateral ventricle's proximity to deep structures, we further showed that LV-BCI supports cognitive BCI tasks. During training, ECoG-implanted rats learned and executed the task more slowly than LV-implanted rats (Supplementary Fig. 2), potentially reflecting interference from cortical immune activation. Using a memory-driven T-maze paradigm and a MSSC, we decoded future spatial choices in rats, achieving accuracies up to 98% with a filter-bank strategy—significantly surpassing ECoG. This suggests that ventricular recordings capture richer pre-decisional information, plausibly from hippocampal-diencephalic circuits. Prior work indicates that hippocampal theta compresses spatial trajectories in time[34], supporting prospective planning; our findings align with this view.

This study has limitations. First, despite the minimally invasive delivery, the implantation pathway still requires penetrating cortical tissue to enter the ventricle; such a tract inevitably causes focal trauma and transient inflammation. Although our histology indicates that microglial activation in the LV group subsides to control levels, the initial injury along the cannulation path is currently unavoidable, and its cumulative impact in repeated accesses or in disease states remains unknown. Second, our work did not directly quantify the device's effect on CSF dynamics. The lateral ventricle participates in CSF circulation and pressure regulation; an expanded electrode could, in principle, perturb flow patterns, pulsatility, or pressure gradients. We did not observe hydrocephalus, ventricular enlargement, or behavioral signs of intracranial hypertension within the study window, but subtle changes in flow or solute transport cannot be excluded without dedicated measurements (e.g., phase-contrast MRI or tracer-based flow assays). Long-term effects on ependymal integrity and ciliary function also warrant investigation. Third, channel counts and spatial coverage are limited at present, largely constrained by the small size of the rat lateral ventricle, which restricts array footprint, routing density, and connectorization. This limits spatial sampling and the ability to perform high-resolution source separation relative to state-of-the-art cortical grids or penetrating arrays.

Future work can mitigate these limitations. To reduce tract trauma, smaller-diameter introducers, soft robotic steerable catheters, and translaminar deployment under endoscopic visualization may help minimize parenchymal disruption. To assess and preserve CSF physiology, integrating in situ pressure sensing, designing lower-profile frames with fenestrations to maintain flow, and performing longitudinal CSF-flow imaging will be important. To expand channel count and coverage, higher-density thin-film interconnects, multiplexed ASICs, and distributed modular "lantern segments" could increase spatial sampling while respecting ventricular constraints. Hybrid systems combining ventricular arrays with cortical ECoG or depth leads may offer complementary bandwidth and depth sensitivity. Finally, integrating stimulation (electrical, optical) and microfluidics [35-37] could enable closed-loop neuromodulation while monitoring CSF biomarkers, creating an avenue for both neuroscience and therapeutic applications.

In sum, we validate the lateral ventricle as a viable BCI access route. With a compliant lantern-inspired electrode, we achieved long-term, wideband recordings and decoded complex memory-related decisions from within the ventricle. This approach offers a design path for stable BCIs and a tool for systems neuroscience. As flexible materials and microsystems mature, passive ventricular recorders could evolve into active interfaces for high-dimensional, multi-scale mapping and intervention. We anticipate that LV-BCI will help bridge information flow between cortex and deep structures, inform therapies for neurocognitive disorders, and expand the scope of brain-computer interaction. By "looking from the ventricle," deeper circuits enter view, advancing our understanding of the neural basis of consciousness and cognition.

## Methods

**Animals**

To evaluate the efficacy and biosafety of the proposed approach, we used 8-week-old male Sprague-Dawley rats. All procedures were approved by the Institutional Animal Care and Use Committee (IACUC) of Tsinghua University (protocol 24-GXR1). Ten rats underwent electrode implantation: five received the lantern-inspired LV electrodes and five received ECoG electrodes. For immunohistochemistry, an additional 27 rats were used: nine with LV electrodes, nine with ECoG electrodes, and nine sham (no implant). One additional rat was used to assess consistency of auditory recordings.

**Electrode fabrication and implantation**

We combined fPCB technology with origami-inspired assembly (workflow in Supplementary Fig. 3). Layouts were designed in Altium Designer and fabricated on polyimide substrates using contact photolithography, achieving a minimum linewidth of 50 μm. After planar array fabrication, ultraviolet laser cutting defined fold lines and outlines. The structures were then reconfigured into a three-dimensional lantern geometry following traditional origami folding. DOWSIL 3145 RTV adhesive (MIL-A-46146) was used to bond the two ends under UV curing, ensuring long-term structural integrity under CSF flow.

To improve electrochemical performance and biocompatibility, a 26 nm gold layer was sputter-deposited at the contact sites. The fPCB was then soldered to a rigid printed circuit board (PCB) using solder paste, and two surface-mount pin headers were attached. Silver wires (0.15 mm diameter) were soldered to the reference and ground pads. Two 3 mm holes at the PCB edge enabled skull fixation.

Electrode delivery referenced an EVD approach under stereotaxic guidance. A small cranial opening was created at a site directly superior to the right lateral ventricle, and the introducer was advanced along a downward trajectory toward the ventricular target. The folded lantern electrode was then deployed within the ventricle under stereotaxic control and expanded in situ. This approach minimized cortical exposure and parenchymal disruption.

Signals were first routed through the PCB, which was fixed to the skull with dental cement and screws, then connected via leads to a NEUSEN W wireless amplifier. Data were transmitted wirelessly to a host computer. The recording system provided 24-bit ADC resolution, 120 dB common-mode rejection, and sub-microvolt input noise (<0.4 μVrms), with a sampling rate of 1 kHz.

**ERP experiments**

SSVEPs are a subtype of event-related potentials[38] with robust expression across rodents and primates [39,40] and are widely used in BCI due to their frequency locking and microvolt-level amplitudes[2,41]. Rats were maintained under gas anesthesia. A custom fixation apparatus aligned the eyes orthogonally to the stimulus monitor, and a black shroud minimized ambient light; retinal responses to photic stimulation persist under anesthesia in rodents.

Each SSVEP trial followed the timeline in Fig. 2d: a 0.5 s baseline (blank screen), a 2 s stimulation period (full-screen target), and a 0.5 s post-stimulus blank. We used a stepwise frequency set spanning 5-20 Hz in 1 Hz increments (16 frequencies). Each frequency was presented in 10 randomized trials.

Stimuli were generated using sinusoidal luminance sampling[42,43]:

$$B(f,i) = Round(255 * 0.5 * \left\{1 + sin\left[2\pi f\left(\frac{i}{RefreshRate}\right)\right]\right\}) \tag{1}$$

where $B \in [0, 255]$ is luminance, f is the target frequency (Hz), i is the frame index, and RefreshRate is 144 Hz. This implements zero-order hold sampling of a continuous sinusoid at Δt = 1/144 s, with spectral properties governed by the sampling theorem.

ASSRs are standard auditory evoked potentials in clinical and engineering applications and serve as robust biomarkers[44,45]. Two precision air-conduction speakers were positioned 5 cm lateral to each ear. Modulated square-wave tones at 30, 40, 60, 70, and 80 Hz were delivered to elicit synchronized neural responses. For each frequency, 50 independent stimuli (1 s pulse width) were presented with 1 s inter-stimulus intervals. SSVEP and ASSR timelines are shown in Supplementary Fig. 5.

**Signal processing**

Data preprocessing. For resting-state analyses, we applied a 50 Hz comb filter. For SSVEP and ASSR, data were first segmented using recorded triggers, trials at the same stimulus frequency were averaged, and then a 50 Hz comb filter was applied. Trials with abnormal total energy were rejected using a Z-score criterion. For the memory decoding task, motion artifacts were manually removed using the ICA module in the MNE toolbox[46]. All filtering operations were implemented with MATLAB R2022a's filtfilt().

Maximum effective bandwidth. Following prior work[8,47], we computed channel-wise PSDs using the multitaper method. We defined the 455-495 Hz band (far from line noise and harmonics within 0-500 Hz) as the noise floor. We compared each 10 Hz band's PSD to the noise-floor PSD: if the median power of a 10 Hz band exceeded the upper whisker of the noise distribution (third quartile plus 1.5×IQR), that band was considered informative. Because biopotentials typically follow a 1/f distribution[48], we scanned 10 Hz bands starting at 0.5 Hz until the significance criterion failed (α=0.05); the starting frequency of that band was taken as the maximum effective bandwidth.

Narrowband SNR. We quantified evoked responses with narrowband SNR:

$$SNR_n = 10 * \log_{10}\left[\frac{N(f_{Target})}{\sum_{\Delta f=-F}^{F} N(f_{Target}+\Delta f))}\right] \quad (2)$$

where $SNR_n$ is the narrowband SNR, $f_{Target}$ is the fundamental or harmonic (for SSVEP/ASSR). In SSVEP, due to slight frequency jitter in the experiment setup, $N(f_{Target})$ included the two adjacent bins (1 Hz total bandwidth). $N(f)$ is the spectral power at frequency f, and F=5 is the narrowband width used in this study.

**Immunohistochemistry**

To assess the sustained impact of different implantation routes on the neuroimmune microenvironment, we examined three postoperative time points: 2 weeks (acute phase), 12 weeks (subchronic), and 24 weeks (chronic). Experimental groups included the LV group (n=3 per time point) and the subdural ECoG group (n=3 per time point), each with age-matched sham controls (n=3 per time point) to ensure adequate comparability. Some animals were survivors from earlier work and were re-enrolled with ethical approval.

Rats were perfused transcardially with phosphate-buffered saline (PBS) to clear blood, followed by 4% paraformaldehyde (PFA) for fixation. Brains were removed, post-fixed in 4% PFA for 24 h, and then cryoprotected in 15% and 30% sucrose solutions for 24 h each. Tissues were embedded in OCT and stored at −20°C. Coronal sections (40 μm) were cut using a cryostat (RWD FS800) and collected on slides.

Sections were rinsed in PBS (3×, 3 min each), then blocked at room temperature for 1.5 h in PBS-T containing 5% bovine serum albumin and 0.3% Triton X-100. Primary antibody incubation was performed overnight at 4°C with rabbit anti-Iba1 (Wako, Code 019-19741; 1:1000 in blocking buffer). The next day, sections were washed in PBS (5×, 3 min each) and incubated with secondary antibody (Invitrogen A21206, labeled donkey anti-rabbit IgG; 1:1000) for 1 h at room temperature, followed by PBS washes (8×, 3 min each) to minimize background. After air-drying, mounting medium was applied, and slides were imaged with an Olympus VS200 whole-slide scanner for high-resolution acquisition.

**Memory decision T-maze task**

Fig. 4a illustrates the behavioral platform, which follows classical T-maze design principles. A transparent acrylic T-maze was mounted on an elevated platform to leverage rats' aversion to heights and encourage them to remain and move within the maze. Because each rat carried an implanted electrode, conventional foot-shock punishment was avoided to protect the implant; instead, metallic tapping sounds were used as aversive cues. The tapping was produced by striking the stainless-steel fence surrounding the T-maze.

To avoid interfering with ERP recordings, behavioral training was scheduled between electrophysiology sessions. After surgery, rats were given 7 days for recovery, followed by a 7-day habituation period to acclimate to the environment and task. After the first recording session, a 12-day familiarization phase was conducted with rewards only and no punishment. After the second recording, formal training began with combined reward and aversive cues for 12 days. Each rat completed 10 trials per day (five left-cued and five right-cued). Throughout the 7 days preceding formal training and during the formal training itself, daily food rations were reduced by approximately 20% to enhance motivation; body weight was monitored at least every other day, and cumulative weight loss never exceeded 15% for any animal. Final recordings were performed once either three consecutive days exceeded 80% correctness or the 12-day formal training was completed; during final sessions, left/right cues were randomized until each side had 10 correct executions. Chocolate pellets were used as the reward during training.

All procedures were conducted in darkness. In a typical trial (Fig. 4b), an LED in the left arm was illuminated, kept on for about 6 s, and then turned off for about 10 s to ensure that choices were memory-guided rather than driven by immediate cues. Neural signals were recorded for 30 seconds during the subsequent pre-decision interval while the rat remained confined. After the cue period, the door was opened and the rat was allowed to navigate freely. A choice toward the previously illuminated arm was scored as correct; a turn to the opposite arm was scored as incorrect.

**MSSC (MicroState Sequence Classifier)**

We developed a microstate sequence-based classifier to decode left/right target intent from deep EEG signals. Continuous 30 s EEG segments were partitioned with a sliding window of 100 ms (15 ms overlap) to preserve temporal continuity and capture short-lived features. For each window, we computed the mean absolute multi-channel amplitude to form a microstate representation:

$$m_t = \frac{1}{W}\sum_{i=1}^{W} |x_{t,i}| \qquad (3)$$

where $m_t$ is the microstate scalar at window t, W is the window length, and $x_{t,i}$ denotes the multi-channel sample at index i within window t. To extract discriminative patterns, microstates from the training set were clustered into eight classes using K-means. For each cluster k, we quantified condition specificity as:

$$\Delta_k = \left| N_k^{Right} - N_k^{Left} \right| \tag{4}$$

where $N_k^{Right}$ and $N_k^{Left}$ are the counts of microstates from right- and left-cued trials, respectively. The four clusters with the largest $\Delta_k$ were selected as discriminative microstates; all others were encoded as 0 (non-discriminative).

Each trial's continuous microstate sequence was mapped to a discrete label sequence, retaining labels for discriminative clusters and setting others to 0, yielding a sparse symbolic series. This sequence was input to a support vector machine (SVM) with an RBF kernel for nonlinear classification. To mitigate class imbalance, we applied data augmentation by perturbing original sequences during training. Performance was evaluated with leave-one-out cross-validation.

**Filter bank**

To exploit complementary information across frequency bands, we decomposed EEG into multiple bands: Delta (0.5-4 Hz), Theta (4-8 Hz), Alpha (8-13 Hz), Beta (13-30 Hz), Gamma (30-80 Hz), and High Gamma (>80 Hz). Each band was processed independently by the MSSC pipeline to produce band-specific class probability estimates.

Final decisions were obtained via weighted voting. Given that different bands contribute unequally to intention decoding, we used an exponential-decay weighting scheme:

$$w_b = b^a + c \tag{5}$$

where b is the band index, $a = -1.25$ is the decay coefficient, and $c = 0.5$ is a bias term. This assigns relatively lower weights to higher-frequency bands, reflecting their putative role during movement preparation. The fused probability was computed as:

$$P_{final} = \sum_{b=1}^{B} w_b \cdot P_b \quad (6)$$

where B is the number of bands and $P_b$ is the probability vector from band b. The predicted class was taken as:

$$\hat{y} = \arg\max(P_{final}) \quad (7)$$

This ensemble strategy integrates complementary spectral cues and improves single-trial decoding accuracy.

## Data availability

All raw data and behavioral videos will be deposited in a public repository upon acceptance of the manuscript.

## Code availability

Selected processing scripts, including the MSSC implementation, will be released to a public repository upon acceptance.

## Acknowledgements

Authors would like to thank Yuqing Zhao from the Central Academy of Fine Arts for their help in drawing the pictures in this article.

## Author contributions

Y. Sun: Data curation, Writing, Methodology, Formal analysis. Y. Gao: Data curation, Formal analysis, Resources. K. Wang: Methodology, Formal analysis. J. Sun: Formal analysis. Y. Chen: Formal analysis, Writing. Y. Yang: Data curation. T. Zhao: Data curation. H. Zhu: Data curation. R. Liu: Reviewing and editing, Formal analysis. X. Chen: Reviewing and editing, Formal analysis, Resources. B. Lu: Reviewing and editing, Resources. X. Gao: Supervision, Reviewing and editing, Project administration.

## Funding

This work is supported by the Postdoctoral Innovation Talents Support Program under Grant BX20250480; the National Natural Science Foundation of China under Grant U2241208; the Non-profit Central Research Institute Fund of Chinese Academy of Medical Sciences under Grant 2024-JKCS-26; the CAMS Innovation Fund for Medical Sciences under Grant 2025-I2M-TS-10; the Beijing Municipal Natural Science Foundation under Grant No.L223035 & No.L246019; the National Key R&D Program of China under Grant No. 2023YFC3502700; the National Natural Science Foundation of China under Grant No. U22A20355, 82027807; the Grant of Tsinghua University (School of Biomedical Engineering) - United Imaging Healthcare Joint Research Center for Magnetic Resonance Imaging; and the Beijing Municipal Science & Technology Commission under Grant Z231100004823012.

## Competing interests

The authors confirm that there are no known competing financial interests or personal relationships that could be perceived to have influenced the work reported in this paper.

# Supplementary Information of

# Lateral Ventricular Brain-Computer Interface System: Lantern-Inspired Implants for Enhanced Performance and Memory Decoding

Yike Sun[1, 2, +], Yaxuan Gao[3, +], Kewei Wang[1, +], Jingnan Sun[1], Yuzhen Chen[1], Yanan Yang[3], Tianhua Zhao[3], Haochen Zhu[3], Ran Liu[1, *], Xiaogang Chen[4, *], Bai Lu[3, *] and Xiaorong Gao[1, *]

+Y. S., Y. G. and K. W. contributed equally to this work.

1 School of Biomedical Engineering, Tsinghua University, Beijing, 100084, China.

2 School of Biological Science and Medical Engineering, Beihang University, Beijing, 100191, China.

3 School of Pharmaceutical Sciences, Tsinghua University, Beijing, 100084, China.

4 Institute of Biomedical Engineering, Chinese Academy of Medical Sciences and Peking Union Medical College, Tianjin, 300192, China.

*Correspondence address. Tsinghua University, Beijing, 100084, China.

Email: liuran@tsinghua.edu.cn; chenxg@bme.cams.cn; lubailab@gmail.com; gxr-dea@mail.tsinghua.edu.cn.

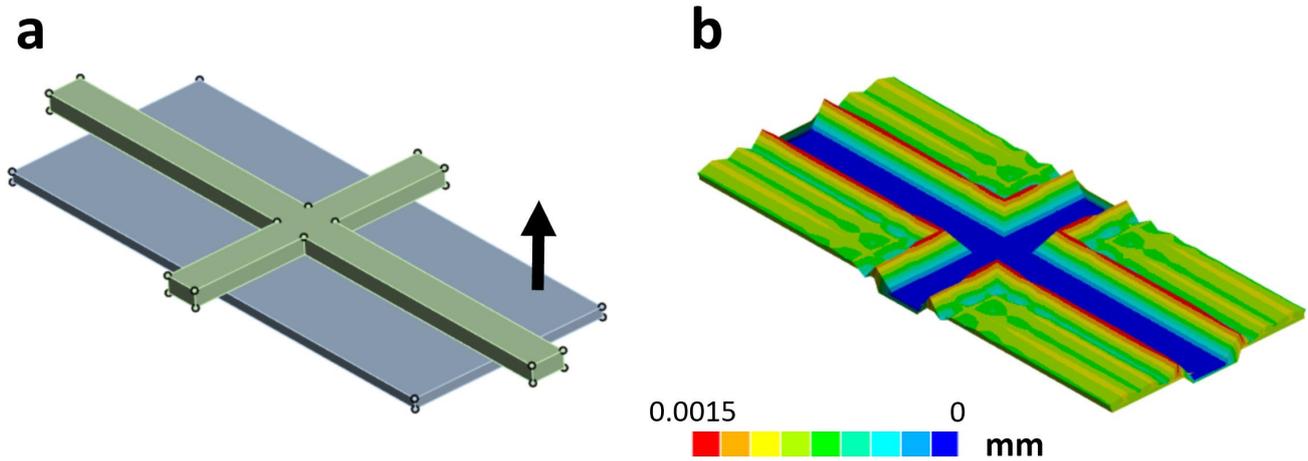

**Supplementary Figure 1.** Simulation of compression between a stent electrode and the ventricular wall. The stent is modeled as a titanium-alloy mesh; the simulation considers only relative displacement between the contact and the ventricular wall. (a) Model setup and direction of relative displacement with a 0.001 mm displacement load. (b) Results showing a peak ventricular wall deformation of 0.0015 mm.

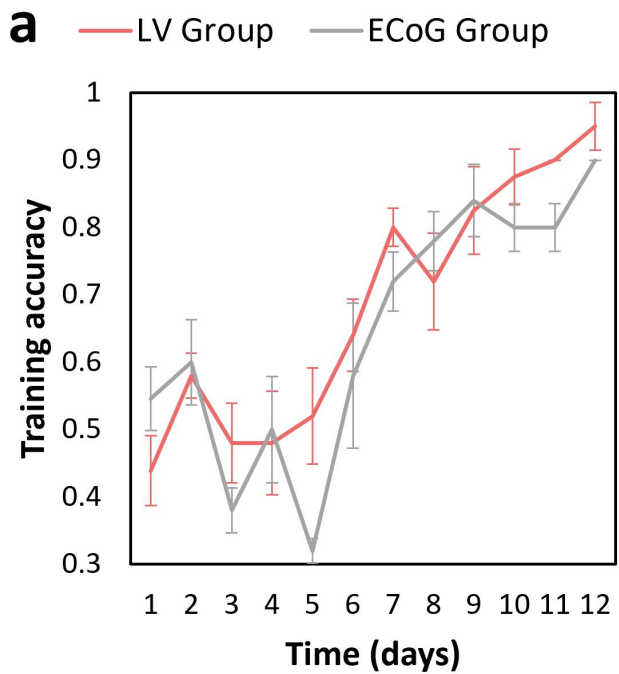 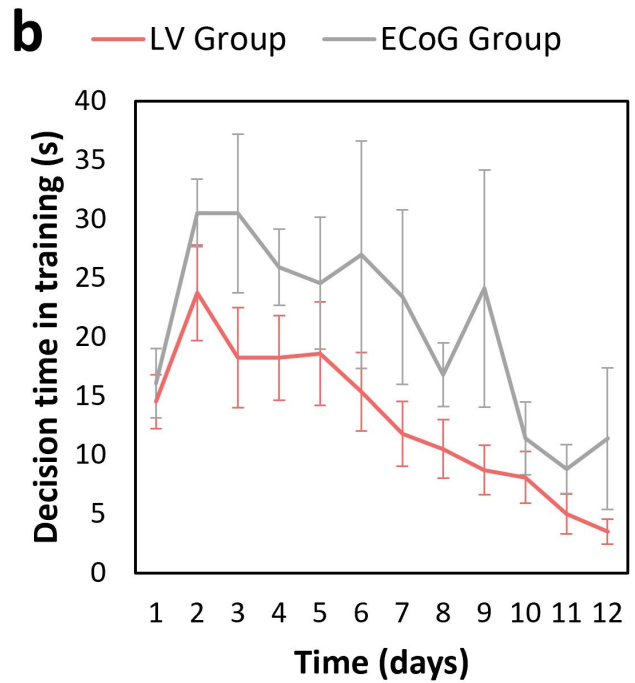

**Supplementary Figure 2.** Training accuracy and decision time during the formal training phase. (a) Evolution of training accuracy; learning rates were similar between the LV and ECoG groups. (b) Changes in decision time; rats in the LV group consistently showed shorter decision times than the ECoG group, potentially due to reduced immune response, though the difference was not significant.

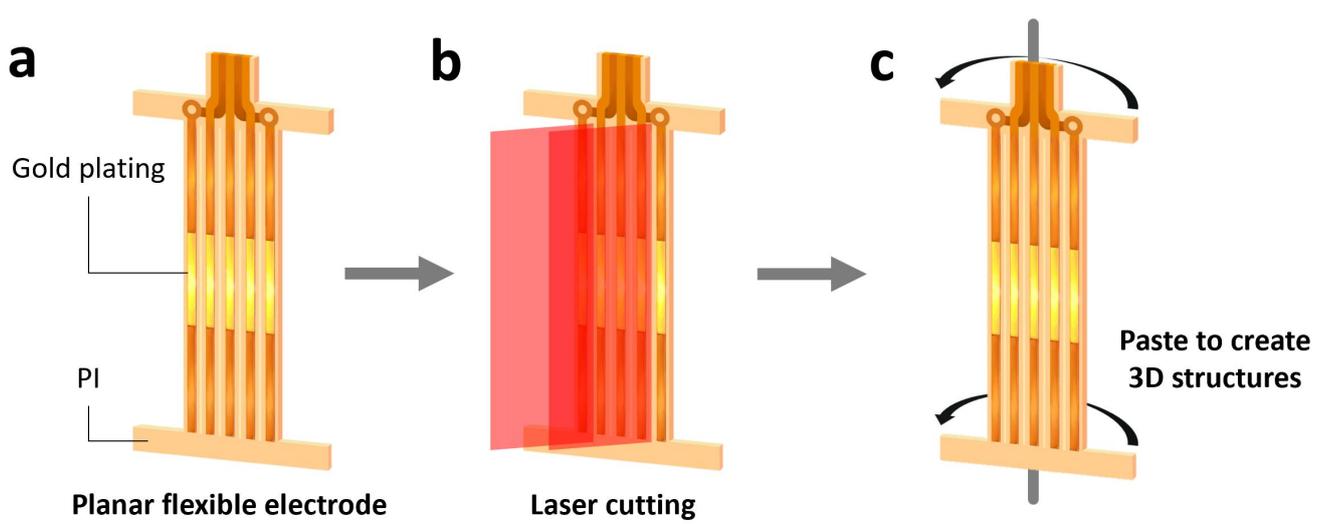

**Supplementary Figure 3.** Workflow for fabricating the lantern-shaped 3D lateral ventricle electrode.

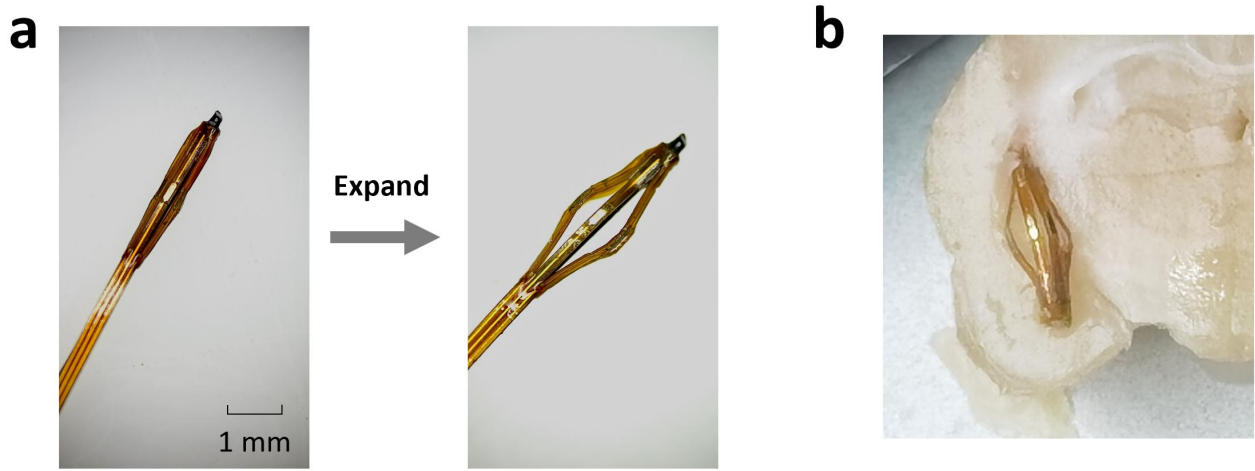

**Supplementary Figure 4.** Deployment sequence and in situ photographs of the lantern-shaped 3D lateral ventricle electrode. (a) Deployment sequence of the electrode. (b) Coronal brain section illustrating the electrode's configuration within the lateral ventricle.

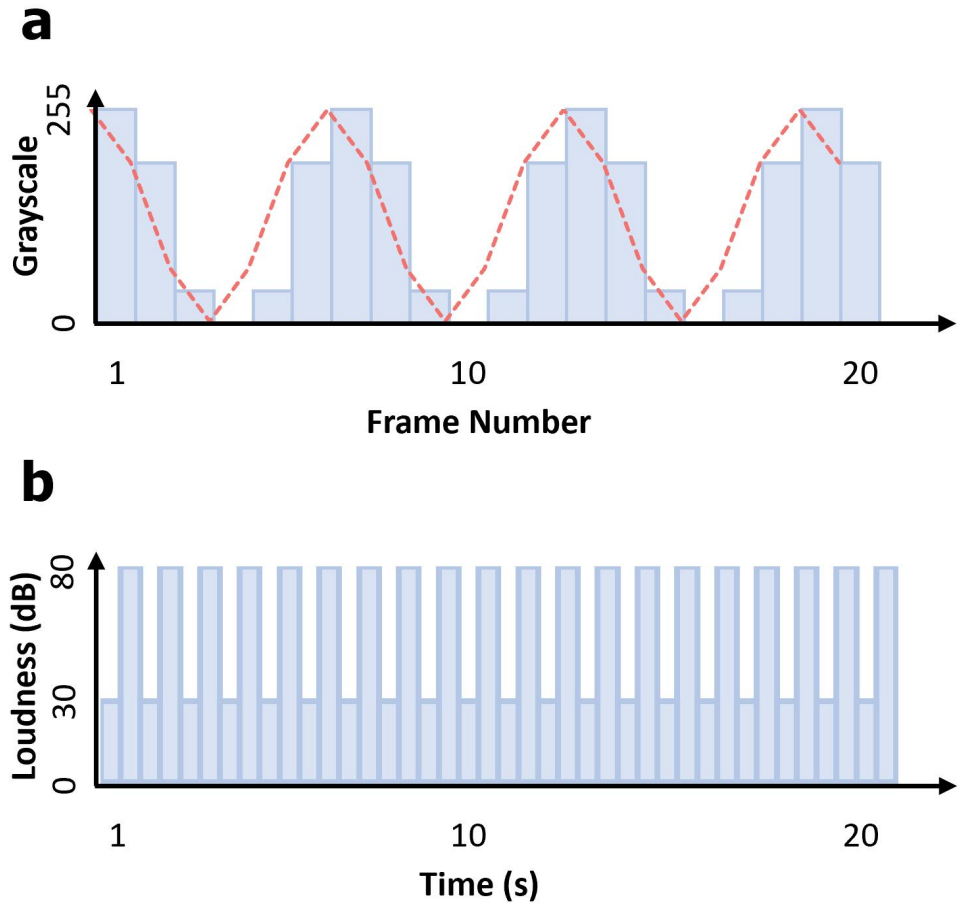

**Supplementary Figure 6.** a. Schematic representation of the SSVEP luminance sequence under 10Hz stimulation. b. Schematic representation of ASSR intensitysequence under 40Hz stimulation.

**Supplementary Table 1.** Parameters for mechanical simulation of the lateral ventricle electrode

| Component | Young's modulus (Pa) | Density (kg m$^{-3}$) | Poisson's ratio |
|---|---|---|---|
| Brain tissue | 10000 | 1050 | 0.45 |
| Titanium alloy | $9.6 \times 10^{10}$ | 4620 | 0.36 |
| Polyimide (PI) | $2.478 \times 10^{9}$ | 1379 | 0.3986 |

**Supplementary Table 2.** Parameters for dynamic simulation of electrode deployment

| Component | Young's modulus (Pa) | Density (kg m$^{-3}$) | Poisson's ratio |
|---|---|---|---|
| Copper interconnect | $1.1 \times 10^{11}$ | 8900 | 0.34 |
| Gold contact | $7.9 \times 10^{10}$ | 19320 | 0.4 |
| Polyimide (PI) | $2.478 \times 10^{9}$ | 1379 | 0.3986 |

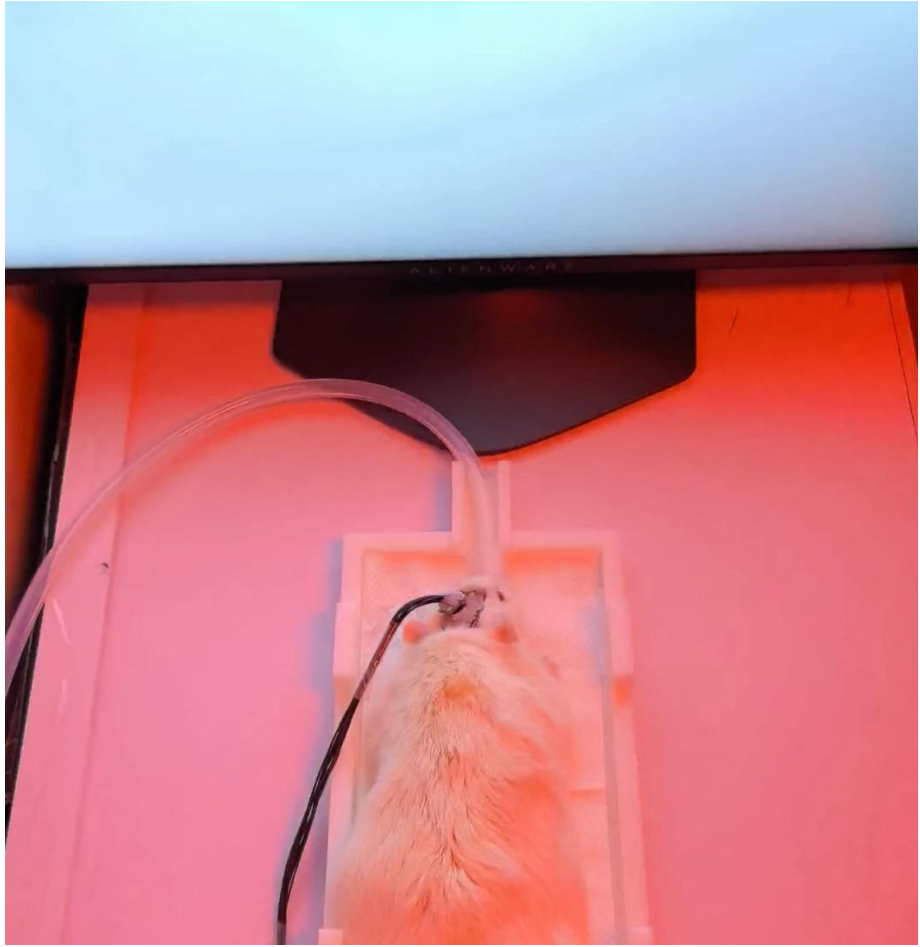

**Supplementary Video 1.** Rat SSVEP experiment. The rat was placed on a flexible resin platform and lightly anesthetized with low-dose isoflurane. Visual stimuli were presented directly in front of the rat. The video contains four visual-stimulation trials. For filming purposes, a red illumination lamp was turned on during recording; actual experiments were conducted in complete darkness.

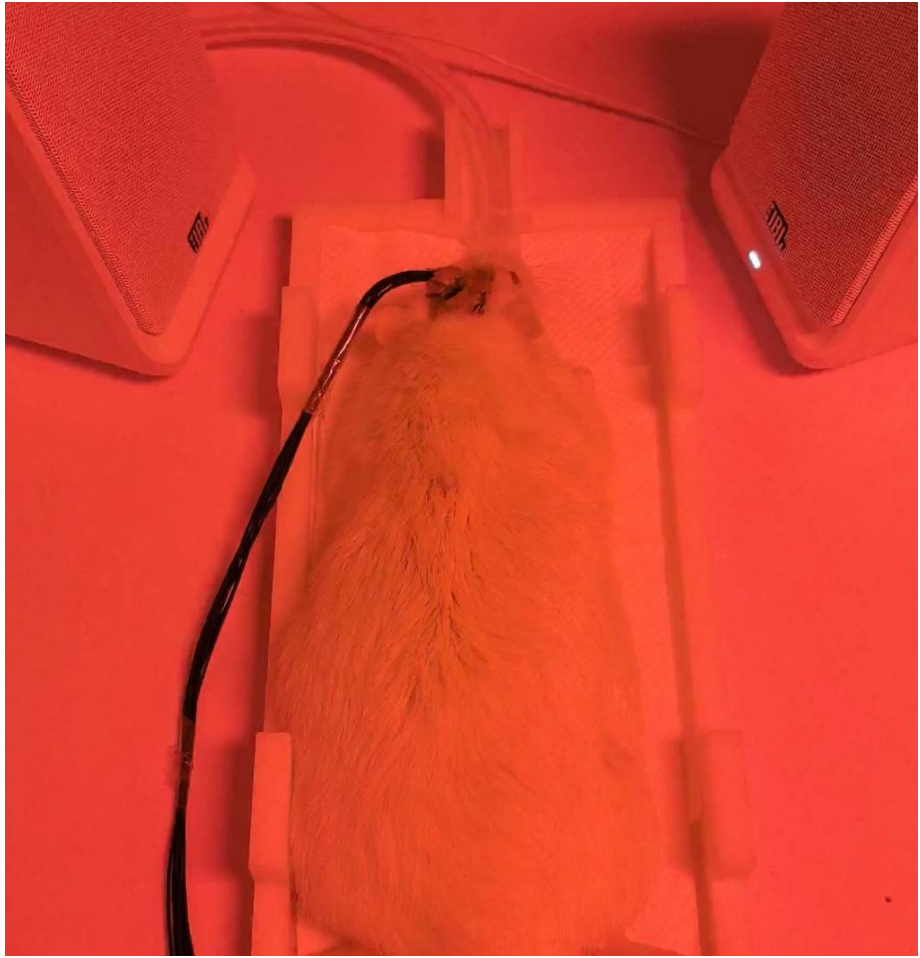

**Supplementary Video 2.** Rat ASSR experiment. The rat was placed on a flexible resin platform and lightly anesthetized with low-dose isoflurane. Pneumatic auditory stimuli were delivered bilaterally. The video contains eight auditory-stimulation trials. For filming purposes, a red illumination lamp was turned on during recording; actual experiments were conducted in complete darkness.

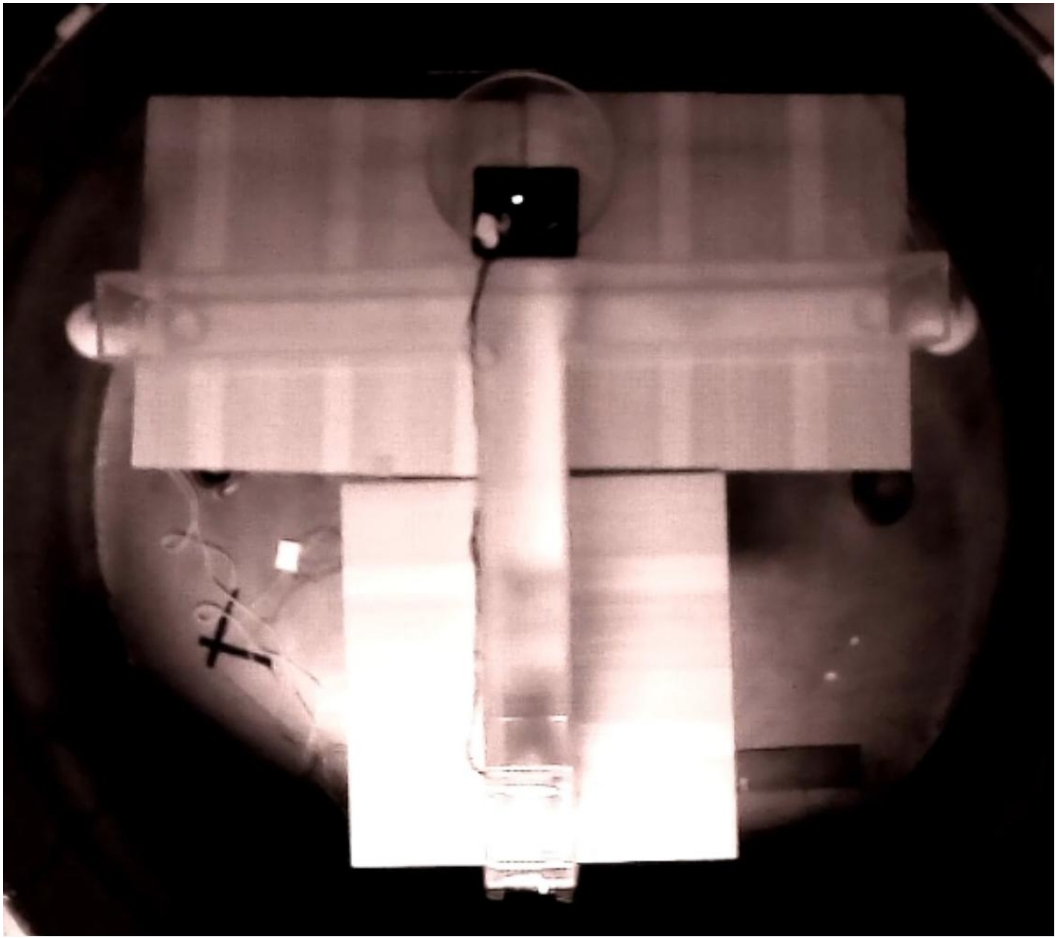

**Supplementary Video 3.** Monitoring video of the behavioral task and electrophysiological recording. The video shows a complete experimental sequence: first, the light in the right arm of the T-maze was turned on for 10 s; next, the light was turned off and neural signals were recorded for 30 s; finally, the rat was placed in the maze and, following training, navigated by memory to the previously illuminated right arm.